\documentclass[twocolumn,showpacs]{revtex4}
\usepackage{citesort} % Zitate sortiert
\usepackage{graphicx} % Abbildungen einfach einbinden
\usepackage[english]{babel} % Schoene Trennung
\usepackage{amsmath} % Viel Mathe
\usepackage{array}

\begin{document}

\title{%$\mathbf{\frac{1}{T}}$
  Thermodynamic Properties of the Dimerised and Frustrated $S=1/2$ Chain}

\author{Alexander B\"uhler}
 \email{ab@thp.uni-koeln.de}
\affiliation{
Institut f\"ur Theoretische Physik, Universit\"at zu
  K\"oln, Z\"ulpicher Stra\ss{}e 77, 50937 K\"oln, Germany}%
\author{Ute L\"ow}
 \email{ul@thp.uni-koeln.de}
\affiliation{
Institut f\"ur Theoretische Physik, Universit\"at zu
  K\"oln, Z\"ulpicher Stra\ss{}e 77, 50937 K\"oln, Germany}%
\author{G\"otz S.~Uhrig}%
 \email{gu@thp.uni-koeln.de}%
 \homepage{http://www.thp.uni-koeln.de/~gu}%
\affiliation{
Institut f\"ur Theoretische Physik, Universit\"at zu
  K\"oln, Z\"ulpicher Stra\ss{}e 77, 50937 K\"oln, Germany}%

\date{\today}

\begin{abstract}
By high temperature series expansion, exact diagonalisation and
temperature density-matrix renormalisation
the magnetic susceptibility $\chi(T)$ and the specific heat
$C(T)$ of dimerised and frustrated $S=1/2$ chains are computed.
All three methods yield reliable results, in particular for
not too small temperatures or not too small gaps. The series
expansion results are provided in the form of polynomials allowing
very fast and convenient fits in data analysis using algebraic programmes.
We discuss the difficulty to extract more than two coupling constants
from the temperature dependence of $\chi(T)$.
\end{abstract}

%05.10.-a Computational methods in statistical physics and nonlinear dynamics
%75.40.Cx Static properties (order parameter, static susceptibility,
% heat capacities, critical exponents, etc.)
%75.10.Jm Quantised spin models
%75.50.Ee Antiferromagnetics
% Noch eine fuer exakte Resultate/Berechnungen

\maketitle

\noindent
{\it Dedicated to Professor E.~M\"uller-Hartmann on occasion of
 his 60$^{\rm th}$ birthday.}
\section{Introduction}
Quantum spin systems are amongst the most interesting and 
challenging problems of
 many-body theory in solid state physics.
Due to their intrinsic many-body quantum
 character it is not possible to compute
even simple quantities like magnetic
susceptibilities or specific heats
in a  straightforward fashion.
But there are by now a number of powerful approaches
like exact diagonalisaton, quantum Monte Carlo, 
temperature density-matrix renormalisation or high temperature
series expansion which yield the desired quantities. 

The first aim of our present paper is to provide high temperature
series data which in combination with some
$T=0$ information can serve as an input for quick data analysis.
High order series expansions constitute an efficient, frequently
used technique \cite{gelfa00}. 
Exact diagonalisation and temperature density-matrix renormalisation
will serve as benchmarks to assess the reliability of the method proposed.
Similar analyses are carried out in Refs.~\onlinecite{johns00a,johns00b} for
unfrustrated dimerised spin chains and ordinary spin ladders.

The second aim of our work is to demonstrate that it is essentially
impossible to deduce from one quantity like the magnetic susceptibility
$\chi(T)$ at not too small temperatures alone more than two
of the three magnetic couplings of dimerised and frustrated spin chains.
 This should caution anybody who is analysing
such data in great detail. To illustrate the type of problem one
can run in we refer the reader to the analysis of (VO)$_2$P$_2$O$_7$
which was analysed in the very beginning as dimerised chain \cite{johns87}.
Then it was thought to be a two-leg spin-ladder with $J_{||}\approx J_\perp$
\cite{eccle94}. But lately unambiguous evidence from inelastic neutron
scattering was found \cite{garre97a} 
that it is a set of weakly coupled dimerised chains
\cite{proko98,uhrig98c}. The magnetic
susceptibility $\chi(T)$ is compatible with both scenarios \cite{barne94}.

\begin{figure}[htbp]
  \begin{center}
    \setlength{\unitlength}{1cm}
    \begin{picture}(8,3)
      \put(-0.5,0){\includegraphics[width=\columnwidth]{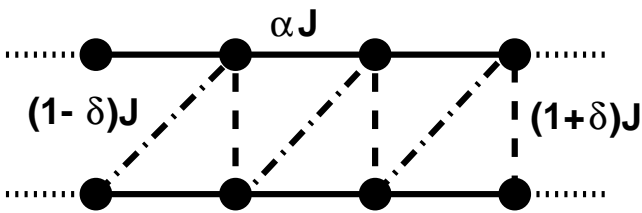}}
    \end{picture}
    \caption{Dimerised and frustrated $S=1/2$ spin chain. For $\delta=1$ a
      two-leg ladder is obtained.}
    \label{fig:dimfrustHM}
  \end{center}
\end{figure}

In this article we focus on gapped one-dimensional spin systems.
These form a rather large class comprising dimerised spin chains,
 strongly frustrated spin chains  but also spin ladders, cf.\
Fig.~\ref{fig:dimfrustHM}.  A representative for a moderately dimerised spin
chain is (VO)$_2$P$_2$O$_7$ \cite{garre97a,proko98,uhrig98c};
a strongly dimerised spin chain realised in
Cu$_2$(1,4-C$_5$H$_{12}$N$_2$)$_2$Cl$_4$
\cite{chabo97a,chabo97b,hamma98,elstn98}; an example for a significantly
frustrated spin chain is the spin-Peierls substance CuGeO$_3$
(see e.g.\ Ref.~\onlinecite{knett01a} and the discussion therein)
which is undimerised in its high temperature phase ($T>T_C\approx14$K) but
weakly dimerised in its low temperature phase.
An important spin ladder compound is SrCu$_2$O$_3$ \cite{azuma94}.

All these substances are interesting because they constitute disordered
antiferromagnets with low coordination number. This means that their
ground state is not given by a N\'eel-type state (i.e.\ with
finite sublattice magnetisation) but by a
Resonating-Valence Bond (RVB) state made of superposed singlet-product
states \cite{liang88}.
If the systems are indeed gapped the average
range of the singlet pairs present
in the ground state is finite. In other words, the correlation length
is finite. Generally, an RVB state is favoured over a N\'eel state
by low coordination numbers, by low values of the spin and by frustration
(which simply weakens the classical ordered N\'eel-state).

If it is possible to dope the insulating magnetic systems unusual
electronic properties emerge due to the strong interplay between
charge and spin degrees of freedom. Some spin ladders like
Sr$_{0.4}$Ca$_{13.6}$Cu$_{24}$O$_{41.84}$ become even superconducting
under pressure \cite{uehar96}. Of course, the appearance of a true phase
 transition at finite temperatures requires a higher dimensionality
than one \cite{mermi66,hohen67,nagat98}. But the driving mechanisms can be
present already in the low-dimensional systems.
So, a deeper understanding of unusual electronic behaviour in doped
antiferromagnets requires a thorough understanding of the magnetic subsystem.
It is in this context that we perform the present investigation
which is designed to determine the relevant magnetic couplings
easily and reliably.

Starting point of our theoretical study is the
Hamilton
\begin{equation}
  \label{eq:hamilton}
  H=\sum_{i=1}^{N}\left(J\left[\left(1+(-1)^i\delta\right)S_iS_{i+1}
  +\alpha S_iS_{i+2}\right] \right)
\end{equation}
with dimerised nearest and and uniform next-nearest neighbour
interaction. The dimerisation is parameterised by $\delta$. The ratio
of nearest and next-nearest neighbor interaction is given by
$\alpha$.  The dimerisation can arise from chemically different bonds
as is the case in (VO)$_2$P$_2$O$_7$.
Alternatively, it may be induced  by a static lattice distortion
via spin-phonon coupling as in CuGeO$_3$. The Hamiltionian can also
be viewed as a spin ladder with an extra diagonal coupling
$(1-\delta)J$ (see Fig.~\ref{fig:dimfrustHM}). In the limit $\delta=1$
it is equivalent to a regular ladder model.  In the limit
$\delta=1$, $\alpha=0$ a system of isolated dimers is obtained.
The ground state properties of the model (\ref{eq:hamilton}) are
investigated recently in numerous papers, see e.g.\
\cite{chitr95,uhrig96b,uhrig96be,white96,brehm96,yokoy97,brehm98,knett00a}.

 In the next section \ref{sec:Methods} we discuss briefly aspects
of the methods we employ. The subsequent section \ref{sec:Chi} is
devoted to the results for the magnetic susceptibility $\chi(T)$.
In particular, we will discuss to which extent several couplings
can be deduced reliably from  $\chi(T)$ data alone.
Section \ref{sec:C} contains the results for the specific heat.
We discuss how much can be learnt from $C(T)$ data. The
Summary \ref{sec:Summary} concludes this article.

\section{Methods}
\label{sec:Methods}
The present work is concerned with the finite temperature properties
of the magnetic susceptibility $\chi(T)$ and the magnetic specific heat
$C(T)$. The analytic method we use is high temperature series expansion (HTSE).
Its results are obtained as polynomials in the coupling parameters with
fractions of integers as coefficients so that no accuracy is lost.
Details of the calculation are found in Ref.~\onlinecite{buhle00}.
Furthermore, the data is provided in electronic form so that they can be
put to use quickly. In order to maximise the range of applicability
some extrapolation schemes are necessary which are described below in
detail.

In addition, we use numerical methods to cross-check the validity
of the analytic results and to supplement them where necessary.
A statistical method is quantum Monte Carlo (QMC) \cite{evert93}.
 In its present form,
however, it suffers from the sign problem for frustrated systems so that
we use it only in the  unfrustrated cases. Mostly we employ
the method of
exact diagonalisation (ED) for smaller finite clusters
\cite{bonne66,fabri98a}. It is based on the
determination of all eigen-values so that the partition sum can be
computed directly. Thereby any thermodynamic
quantity is accessible. For not too low temperatures the results
describe reliably the thermodynamic limit. 
The third technique that we  use is temperature
density-matrix renormalisation (T-DMRG)\cite{klump99a,johns00b}. By this
numerical method
very low temperatures can be reached reliably.

We deal with the static magnetic susceptibility
\begin{equation}
  \label{eq:susceptibility}
  \chi(\beta;\delta,\alpha) =
  \frac{\beta}{N}\frac{\mathrm{tr}M^2e^{-\beta
  H}}{\mathrm{tr}e^{-\beta H}} = \frac{\beta}{N}\langle M^2 \rangle
\end{equation}
and the specific heat
\begin{equation}
  \label{eq:specificheat}
  C(\beta;\delta,\alpha) = \frac{1}{N}\frac{\partial}{\partial T}
  \left( \frac{-\frac{\partial}{\partial \beta}\mathrm{tr}e^{-\beta H}}{
  \mathrm{tr} e^{-\beta H}}\right)
\end{equation}
In the Appendices the series coefficients are given. For the
dimerised and frustrated chain the coefficients for both quantities
$\chi(T)$ and $C(T)$ are
provided up to order 10 in the inverse temperature $\beta$.
For the unfrustrated dimerised chain they are given up to order 18.
The bare truncated series, however,
are not sufficient to describe the  quantities under study at low values of
$T$. To enhance the region of
validity of the high temperature series expansion
Dlog-Pad\'e representations \cite{domb89} are used. We show in the following
that the complete susceptibility of gapped spin chains can be obtained if
some low temperature information is incorporated in a simple way.
This is explained subsequently in detail.

A Dlog-Pad\'e approximant of the expansion of the  magnetic
susceptibility $\chi(\beta)$ is given by
\begin{equation}
  \label{eq:chidlog}
  \chi(\beta) = \frac{1}{4T}exp\left(\int\limits_0^{\beta}P^k_l(\beta
  ')d\!\beta '\right) \ ,
\end{equation}
where $P^k_l(\beta')$ is the rational Dlog-Pad\'e approximant with
a polynomial of degree $k$ in the numerator and a polynomial of
degree $l$ in the denominator of
\begin{equation}
  \label{eq:chipadeln}
  f =\partial_{\beta '}\ln (4T\chi(\beta ')) =
  \frac{\partial_{\beta '}4T\chi}{4T\chi} \ .
\end{equation}
Possible orders $[k,l]$ of $P_l^k(\beta)$ have to fulfill
$k+l=n - 1$ where $n$ is the order of the truncated series available.
 This can be seen by
comparing the number of series coefficients obtained and the number of
 independent (one less than the total number)
 coefficients in the two polynomials of $P_l^k(\beta)$.
 Due to the derivative on the right side of (\ref{eq:chipadeln})
 one coefficient of the series expansion is lost.

Additionally, we incorporate $T=0$ information to supplement the
high temperature expansion. In the case of ungapped spin chains
\cite{buhle00} we used knowledge of the full dispersion for this
purpose. Though successful this approach is cumbersome in general.
Therefore, we now simplify the procedure to incorporate
$T=0$ information considerably. It is described for the case
of gapped spin chains.

At zero temperature the susceptibility of a gapped system vanishes and
at finite but small temperature the deviation is exponentially small
due to the spin gap $\Delta$
\begin{equation}
  \label{eq:chilowT1}
  \chi(T) \approx e^{-\frac{\Delta}{T}}\ \text{ for T}\ll \Delta \ .
\end{equation}
Furthermore, the leading power in $T$ can be determined on the basis of the
dimensionality of the problem and of the
behaviour of the dispersion close to its minima. For one-dimensional
systems with quadratic minima , which is generic for gapped systems,
one obtains \cite{troye94}
\begin{equation}
  \label{eq:chilowT2}
  \chi(T) \approx \frac{1}{\sqrt{T}}e^{-\frac{\Delta}{T}}\ \text{ for T}\ll
\Delta \ .
\end{equation}
This equation provides information for two additional coefficients of
the  Dlog-Pad\'e approximant. Let us extend the series expansion
of $\chi$ by  two  terms $a\cdot \beta^{n+1} +
b\cdot \beta^{n+2}$ yielding $\chi_{\rm e}$. Then the degree of the
approximant $P_l^k(\beta)$ can be incremented by two fulfilling $k+l=n+1$.
The two additional conditions deriving from (\ref{eq:chilowT2})
follow from
\begin{equation}
  \label{eq:betainf}
  P_l^k(\beta) = -\Delta
  -1/(2\beta)+\mathcal{O}(\beta^{-2})
\end{equation}
in the limit $\beta\to \infty$. This condition requires that $P_l^k(\beta)$
is finite for $\beta\to \infty$ imposing the severe constraint $l=k$
on the possible degrees of the Dlog-Pad\'e approximant. To circumvent this
 constraint we substitute
\begin{equation}
  \label{eq:substitut}
  u =\beta/(1+\beta) \Leftrightarrow \beta =u/(1-u)
\end{equation}
thereby mapping the complete $\beta$-interval $[0,\infty]$
to the $u$-interval $[0,1]$. This mapping is justified by the continuity
of $P_l^k(\beta)$  in the limit $\beta\rightarrow\infty$.

The asymptotic behaviour (\ref{eq:betainf}) transforms under the
mapping (\ref{eq:substitut}) to
\begin{subequations}
  \label{eq:detparameter}
  \begin{eqnarray}
    P^k_l(u)\Big|_{u=1} &=& -\Delta \\
    \partial_u P^k_l(u)\Big|_{u=1} &=& 1/2\ ,
  \end{eqnarray}
\end{subequations}
where we use $P^k_l$ now for the rational function in $u$. We will henceforth
consider only approximants in $u$ so that no confusion should arise.

The Dlog-Pad\'e approximant $P_l^k(u)$ is chosen such that it
approximates $\partial_{u}4T\chi_{\rm e}/4T\chi_{\rm e}$
where $\beta$ is expressed
according to (\ref{eq:substitut}) as function of $u$.
In this way, reliable interpolations between the low
temperature behaviour and the high temperature series expansion can be
obtained for arbitrary orders $k$ and $l$ complying with $k+l=n+1$.

To assess the range of validity of the Dlog-Pad\'e approximant
various orders  of $P^k_l$ are investigated. Examination of the positions
$\beta_i$ of the poles of $P^k_l$ and especially of the largest modulus
of them leads to a reliable estimate of the radius of convergence
$\beta_{\rm r}$ of the series in $\beta$ (Cauchy's theorem).
We find roughly $\beta_{\rm r}\approx 2/J$ as
for the ungapped spin chains \cite{buhle00}. This means that the
truncated series will {\it always} diverge around $T\approx 0.5J$ even for
arbitrary high order in $\beta$.
Thus the polynomial representation is not sufficient to describe the maximum
 of $\chi(T)$ quantitatively.

\section{Magnetic Susceptibility $\chi(T)$}
\label{sec:Chi}
\subsection{Applicability of the High Temperature Series Expansion}
\begin{figure}[htbp]
  \begin{center}
    \leavevmode
    \includegraphics[angle=-90,width=\columnwidth]{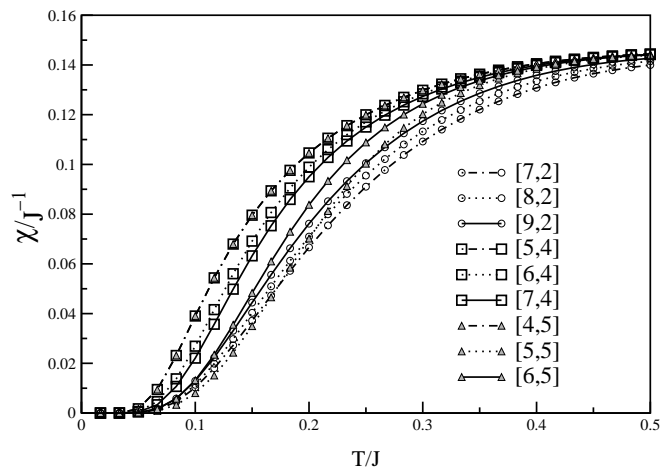}
    \caption{Susceptibility for $\delta=0.1$ and $\alpha=0.24$ for the
      dimerised, frustrated spin-1/2 chain. Various orders of the
      Dlog-Pad\'e approximant are shown.}
    \label{fig:chidimfrustdlogs}
  \end{center}
\end{figure}

 In Fig. \ref{fig:chidimfrustdlogs}
various representations of the susceptibility are shown and compared.
 For fixed order $l$ of
$P_l^k$ the Dlog-Pad\'e representation of $\chi$ moves for $l=2$
upwards, for $l=4$ downwards and for $l=5$ in both directions (for $l=3$
no evaluation is possible due to defective approximants in all orders). All
representations converge for increasing order.
\begin{figure}[htbp]
  \begin{center}
    \leavevmode
    \includegraphics[angle=-90,width=\columnwidth]{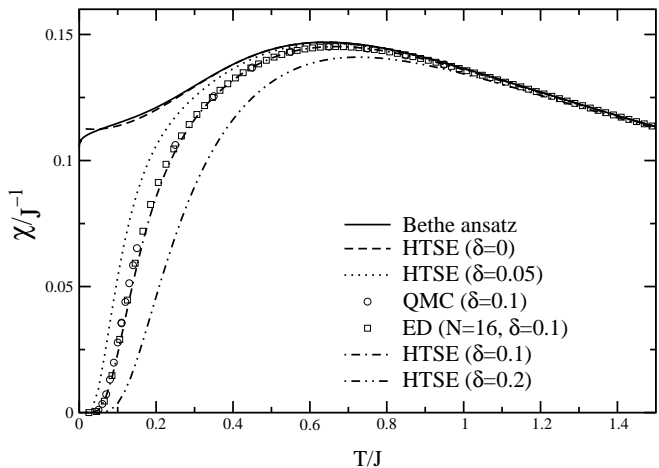}
    \caption{Susceptibility of the dimerised chain for various values of
      $\delta$. The Dlog-Pad\'e approximants are of order $[15,4]$}
    \label{fig:susi_dim_range}
  \end{center}
\end{figure}
The
$[5,4]$ and $[4,5]$ approximants cannot be discerned.
But this is not a general feature. Other reflected approximants
of the type $[k,l]$ and $[l,k]$ yield substantially differing
results. In order to present our results in a systematic and unbiased
way we choose the $P^m_4$ representation for all curves
shown below (if not denoted otherwise).
Deduced from Fig.~\ref{fig:chidimfrustdlogs}
we expect quantitatively reliable results down to $T/J\approx 0.25$ for
the dimerised, frustrated chain. This conclusion is based on 
considering the highest orders and on looking for
 the range of $T/J$ where they are consistent.

In the purely dimerised case, see Fig.~\ref{fig:susi_dim_range}, almost the
whole temperature regime is excellently described. This is due to the high
orders reached ($\mathcal{O}(\beta^{18})$). In Fig.~\ref{fig:susi_dim_range},
 the HTSE results
are depicted  in  comparison to results from numerical methods (ED, QMC)
and the exact result of the uniform chain \cite{klump93b}.
In particular, the agreement between the HTSE result and the exact one
for the uniform chain is impressive.
 We think that this is the optimum
which can be obtained by high temperature expansion since it is certainly
not possible to assess the logarithmic low temperature corrections coming
from the high temperature end. Technically, we used for the uniform chain
instead of (\ref{eq:detparameter}) the obvious relations
\begin{subequations}
\begin{eqnarray}
  \label{eq:detparameter-uniform1}
  P^k_l(u)\Big|_{u=1} &=& 0 \\
  \label{eq:detparameter-uniform2}
  \partial_u P^k_l(u)\Big|_{u=1} &=& 1\ .
\end{eqnarray}
\end{subequations}
The second relation
(\ref{eq:detparameter-uniform2}) reflects the fact that $\chi(0)$ is finite.

\begin{figure}[htb]
  \begin{center}
    \leavevmode
    \includegraphics[angle=-90,width=\columnwidth]{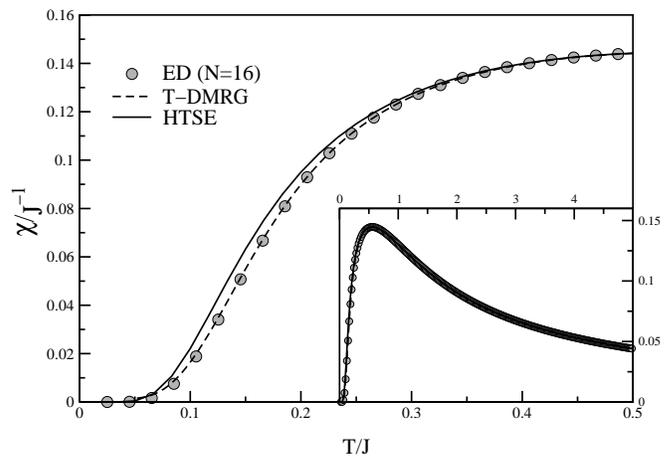}
    \caption{Comparison of high temperature series expansion, exact
      diagonalisation and temperature density-matrix renormalisation data for
      $\alpha=0.24$ and  $\delta=0.1$.}
    \label{fig:chia024d010htse_ed}
  \end{center}
\end{figure}

In Fig.~\ref{fig:chia024d010htse_ed}
the  $[7,4]$ HTSE representation chosen is compared to exact
diagonalisation and  temperature density-matrix renormalisation
data \cite{klump99a}. The results are in very good accordance with
each other. Only in the regime $T/J<0.2$ there is a slight
difference between the HTSE representation and the numerical
results.

\begin{figure}[htb]
  \begin{center}
    \leavevmode
    \includegraphics[angle=-90,width=\columnwidth]{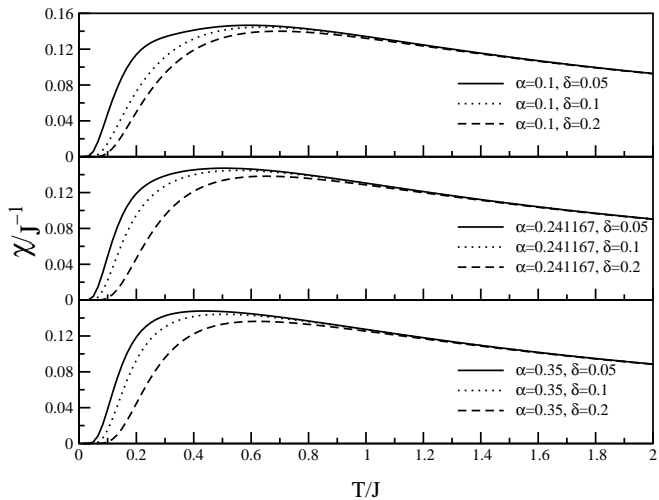}
    \caption{Susceptibility for various values of $\delta$ and
      $\alpha$. Representation by the $[7,4]$ approximant except for
      the solid line in the uppermost panel where the $[7,4]$ approximant
      is defective and the $[8,3]$ approximant is used.}
    \label{fig:chi_htse_many}
  \end{center}
\end{figure}

In Fig.~\ref{fig:chi_htse_many} the susceptibilities for various
sets of parameters are shown. The behaviour of the maximum of the
susceptibility depends upon the parameters under study.
For fixed  next-nearest neighbour interaction $\alpha$
the position of the maximum moves to higher values of $T/J$
for increasing $\delta$ while the maximum value decreases.
These effects are induced by the increasing gap.

Fixing $\delta$, the position of the maximum moves to the
left for increasing $\alpha$. This can be understood from the
reduction of the dispersion on increasing frustration.
The mobility of the excitations is more and more restricted
\cite{uhrig96b,uhrig96be,knett00a}. The maximum  value of $\chi(T)$
 remains almost constant.
This can be seen as the result of an compensation of two contrary
effects. On the one hand, the susceptibility would rise due to the
shift of the maximum position to lower temperatures where the global
$1/T$ factor (cf.~(\ref{eq:susceptibility})) enhances its value.
But on the other hand, the frustration provides an additional
antiferromagnetic coupling in the system which works against an
alignment of the spins. For instance, the antiferromagnetic next-nearest
neighbour coupling induces a strong repulsion between aligned
adjacent triplets on the dimers \cite{uhrig96b,uhrig96be,knett00a}.

\subsection{Information Content of $\chi(T)$}

In this section we address the question to which extent
 the parameters of the model Eq.~\ref{eq:hamilton} can be
extracted from measurements of the susceptibility. In other words,
we adopt the experimentalist's point of view who wants to
determine the coupling parameters from experimental data.
Obviously, the main feature in the susceptibility curve is the
maximum. So it is natural to use in the first place the maximum
value $\chi_{\rm max}$ and its position $T=T_{\rm max}$. We
consider the product $\chi_{\rm max} T_{\rm max}$  since it is
experimentally easily accessible and does not depend on the
exchange coupling $J$.

\begin{figure}[htb]
  \begin{center}
    \leavevmode
    \includegraphics[angle=-90,width=\columnwidth]{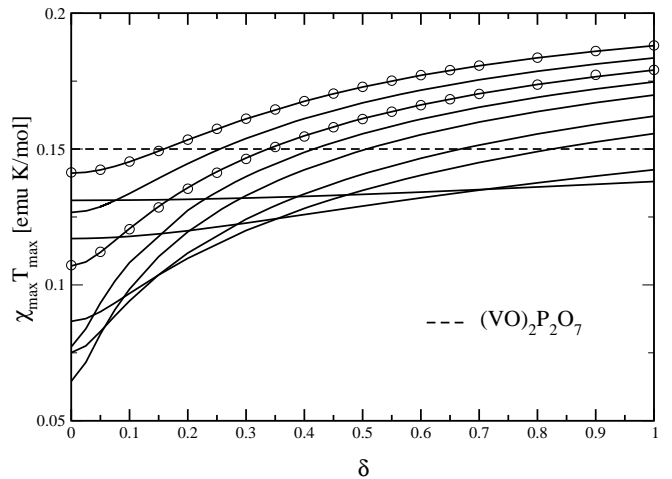}
    \caption{ $\chi_{\rm max} T_{\rm max}$ versus $\delta$ for $\alpha=0.0,
      0.12,0.24,0.36,0.50,0.75,1.0, 2.0$ and $4.0$ (in descending order at
      the right side of the graph, i.e.\ $\delta=1.0$). 
      The symbols are obtained by analysing
      the HTSE showing excellent agreement with the numerical data. For 
      illustration, the dashed
      line refers to the experimental value of (VO)$_2$P$_2$O$_7$.}
    \label{fig1}
  \end{center}
\end{figure}

The data shown in Fig.~\ref{fig1} are obtained by exact
diagonalisation of the full Hamitonian of a 16 site system. For
$\alpha <0.75$ the data are exact to many digits for a system of
this size. For $\alpha >0.75$ and $\delta=0$ finite-size effects
of several percent occur.

Note that for $\delta=0$ the quantity $\chi_{\rm max} T_{\rm max}$
reaches its minimum at $\alpha\approx 0.5$ and then starts to
increase again. This is due to the fact, that on growing $\alpha$
the system approaches two independent chains of half the size of
the original chain with $\alpha=0$.
Fig.~\ref{fig1} can be used easily: given the experimental input
for $\chi_{\rm max} T_{\rm max}$ one can read off the value for
$\delta$ for a chosen $\alpha$.

\begin{figure}[htb]
  \begin{center}
    \leavevmode
    \includegraphics[angle=-90,width=\columnwidth]{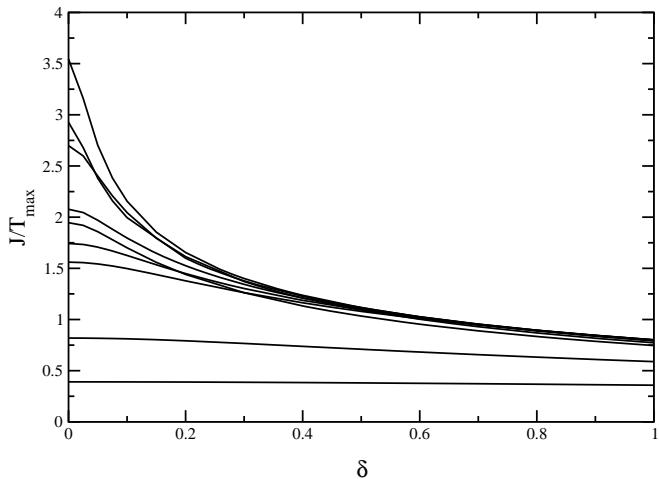}
    \caption{ $\frac{J}{T_{max}}$ versus $\delta$ for $\alpha=0.5,
      0.36,0.75,0.24,1.0,0.12,0.0, 2.0$ and $4.0$ (in descending order at
      the left side i.e. $\delta=0.0$).}
    \label{fig2}
  \end{center}
\end{figure}

To complete the analysis we plot in Fig.~\ref{fig2} the variable
$J/T_{\rm max}$ as a function of $\delta$ for various values of
$\alpha$. So, once the value of $\delta$ (for given $\alpha$) is
determined from Fig.~\ref{fig1}, Fig.~\ref{fig2} helps to
determine the exchange coupling $J$ by reading off $J/T_{\rm max}$
and multiplying by $T_{\rm max}$.

We want to point out here that it is almost impossible to
determine the values of $J$, $\alpha$ {\em and} $\delta$ from the
temperature dependence of the susceptibility alone (cf.\ also
\onlinecite{low00}). This phenomenon is well known from the
investigations of (VO)$_2$P$_2$O$_7$. In the case of this
substance the susceptibility of the isotropic ladder i.e.\ a
ladder with $J_{||}=J_\perp$ and the susceptibility of a dimerised
spin chain with $\delta=0.2$ fit both the experimental data
equally well. For illustration the possible  results
 for (VO)$_2$P$_2$O$_7$, where $\chi_{\rm max}=2.07\ 10^{-3}$ emu/mol V
and  $T_{\rm max}=74K$ correspond to the horizontal dashed line at
0.15 emu K/mol V in Fig.~\ref{fig1}.
\begin{figure}[htb]
  \begin{center}
    \leavevmode
    \includegraphics[angle=-90,width=\columnwidth]{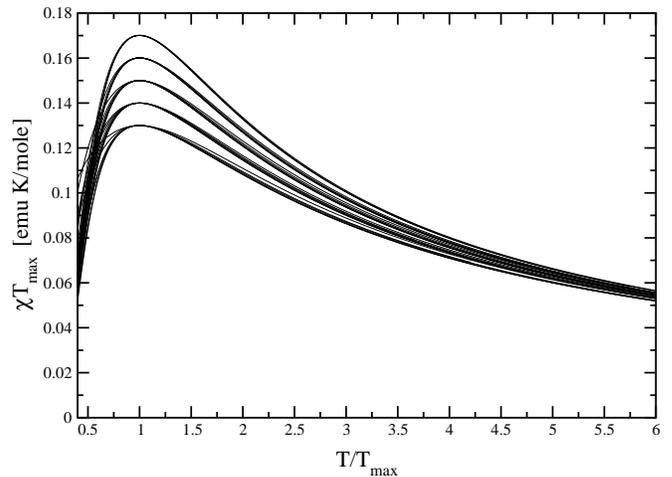}
    \caption{Re-scaled susceptibilities  for
        $\chi_{\rm max} T_{\rm max}$= 0.17, 0.16, 0.15, 0.14, 0.13 emu K/mol
        for the $\alpha$ and $\delta$ values shown in
        Fig.~\protect{~\ref{fig4}} ($g$-factor set to 2).}
    \label{fig3}
  \end{center}
\end{figure}
The difficulty to distinguish different sets of ($J, \delta,
\alpha$) yielding the same value of $\chi_{\rm max} T_{\rm max}$
is visualised strikingly in Fig.~\ref{fig3}. The re-scaled
susceptibilities belonging to various values of $\chi_{\rm max}
T_{\rm max}$ are depicted. In the temperature region around the
maxima and for larger temperatures the differences within each set
are minute. Thereby we conclude that it is impossible to determine
all three coupling parameters from $\chi(T)$ at moderate and at large
values of temperature alone. In Fig.~\ref{fig4} the values of
$\alpha$ and $\delta$ are shown which belong to the various sets
displayed in Fig.~\ref{fig3}.
\begin{figure}[htb]
  \begin{center}
    \leavevmode
    \includegraphics[angle=-90,width=\columnwidth]{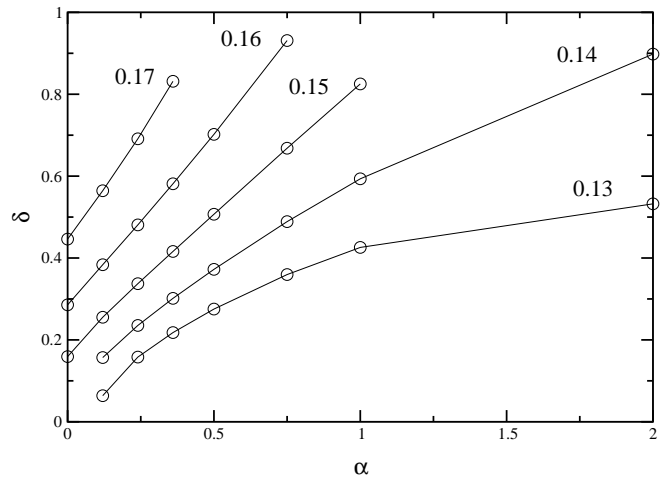}
    \caption{Lines of constant height  $\chi_{\rm max} T_{\rm max}$=
      0.17,0.16,0.15,0.14,0.13 emu K/mole in the  $\alpha$-$\delta$ plane.}
    \label{fig4}
  \end{center}
\end{figure}

Next we use the high temperature expansion results (see Appendix A2)
to understand
such a scaling behaviour in the lowest orders of $\beta$. It is
obvious that the zeroth order $4 T \chi \approx 1$ does not allow
the determination of any parameters. The first order of the
susceptibility of a model with given frustration $\alpha$ and
exchange coupling $J$ is identical to the first order of another model
with frustration $\alpha_1$ and coupling $J_1$ if
\begin{equation}
  J_1(1+\alpha_1)=J(1+\alpha)
  \label{o2a}
\end{equation}
holds. In other words, if one had only first order results for the
susceptibility  it would be impossible to determine $J$ and
$\alpha$ independently.

The second order of the HTSE depends on $\delta$. But again one
can choose a particular value $\delta_1$ such that the sets
$(J, \alpha, \delta)$  and $(J_1, \alpha_1, \delta_1)$ lead to identical
zeroth, first and second order terms in $\beta$. This choice is
\begin{equation}
  \delta_1^2=2 \alpha_1-(2\alpha-
  \delta^2)\left(\frac{1+\alpha_1}{1+\alpha}\right)^2 \ .
  \label{o3}
\end{equation}
\noindent
If the susceptibility is determined mainly by the first three orders
the relations (\ref{o2a},\ref{o3}) provide the recipe to re-scale
the susceptibility such that different parameter sets yield similar
temperature dependences. Indeed, if one focuses on the high temperature range
this is true. The precise position and value of the maxima, however,
cannot be deduced from the first three orders alone. This implies that
Eq.~\ref{o3} provides only rough estimates for the curves displayed in
Fig.~\ref{fig4}.

To be cautious, we like to stress that we are not claiming that it is
impossible to determine all three coupling $J, \alpha, \delta$ {\it if}
sufficient low temperature data is available to obtain the value of the
gap. The gap $\Delta$ depends in a different way on the couplings than 
$\chi(T)$ does \cite{knett00a}. Very often, however,
 the dependence at low values of the temperature
is either not accessible or it no longer corresponds to a pure 1D system.
In particular, interchain couplings $J_\perp$
lead to significantly altered gaps
(for examples see Refs.~\onlinecite{knett01a,uhrig01a}) although the behaviour
at higher temperatures $T> J_\perp$ is still well described by a 1D model.

\section{Specific Heat $C(T)$}
\label{sec:C}
It is a straightforward idea
to extract further information about
the magnetic properties of certain materials by considering
also the specific heat $C$.
Only in very rare cases, however, the  magnetic part of $C$ can be extracted
in a reliable way from the measured data because
the phononic contributions dominate $C$ whenever the energy scale
of the lattice vibrations is of the order of the magnetic coupling $J$.

At low temperatures where the phononic contributions vanish following the
usual $T^3$ law a reliable  extraction of $C$ is possible.
For high temperatures such a procedure will  fail in general
as can be seen, for instance, in a
simple model of Einstein (dispersionless) phonons coupled to
Heisenberg chains \cite{kuhne99b}.
So the full temperature dependence of the magnetic part of $C$
can be measured only in
substances with a small exchange coupling $J$ as
it occurs, for instance, in organic magnetic materials, see e.g.\
Ref.~\onlinecite{rapp95}.

Furthermore, it is in order to mention that there are indirect
techniques to obtain the magnetic part of $C(T)$ where the
energy fluctuations are linked to dissipation. The latter is
measured by the intensity of elastic scattering in spectroscopic
investigations, see e.g.\ \cite{kuroe97,grove00}. The indirect approaches,
however, may provide information on $T_{\rm max}$ of $C(T)$  but not
on $C_{\rm max}$ itself since overall factors are not known.
(In
this section $T_{\rm max}$ refers always to $C(T)$. The position
of the maximum of $\chi(T)$ is denoted $T_{\rm max}^\chi$.)

%\paragraph{HTSE}
In the Appendices A and B the coefficients for the specific heat
are provided. In order to compute $C(T)$ additional information
on the low temperature behaviour is included as was done for
$\chi(T)$. At zero temperature the
magnetic specific heat of a gapped system vanishes and at finite but
small temperatures the deviation  is
exponentially small. For one-dimensional systems one has \cite{troye94}
\begin{equation}
  \label{eq:specificheat_lowT}
  C(T) \propto T^{\frac{3}{2}}e^{-\frac{\Delta}{T}}\ .
\end{equation}
This asymptotic behaviour leads to a Dlog-Pad\'e representation of $C$
equivalent to the one of the susceptibility (see
Eqs.~\ref{eq:chidlog} to \ref{eq:detparameter}). The
function approximated by $P_l^k$ reads
\begin{equation}
  \label{eq:specificheta_padeapp}
  g = \partial_{\beta }\ln \left(\left[\frac{3}{16}\left(1+ \delta^2
  +\alpha^2 \right)\beta^2\right]^{-1} C(\beta)\right)
\end{equation}
where the additional factor ensures that the argument of the
logarithm tends to unity for $\beta\to 0$.
As for the susceptibility two additional terms are taken into account
which are determined such that the correct asymptotics in $u=\beta/(1+\beta)$
is guaranteed
\begin{eqnarray}
  \label{eq:specificheat_param}
  P_l^k(u)\big|_{u=1} &=& -\Delta \\
  \partial_u P_l^k(u)\big|_{u=1} &=&  7/2\ .
\end{eqnarray}

\begin{figure}[htb]
  \begin{center}
    \leavevmode
    \includegraphics[angle=-90,width=\columnwidth]{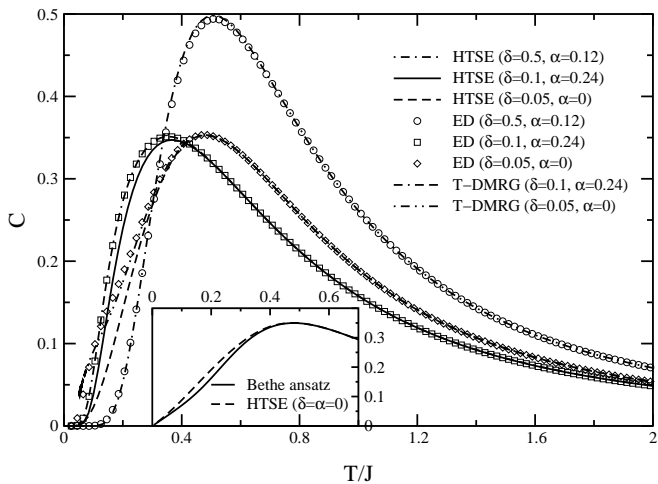}
    \caption{Specific heat $C$ for various values of $\delta$ and
      $\alpha$. The Dlog-Pad\'e approximants used are of degree $[5,4]$ for the
      frustrated and dimerised chains and $[13,4]$ for the chains
      without frustration.}
    \label{fig:heat_comp}
  \end{center}
\end{figure}
In Fig.~\ref{fig:heat_comp} the specific heat for various  sets of
parameters $\alpha$ and $\delta$ is compared to numerical ED and T-DMRG
results. For large enough dimerisation $\delta$, the  ED and HTSE results
agrees perfectly. For $\delta=0.1$ and $\alpha=0.24$ the HTSE describes the
position of the maximum of the specific heat very well,
but the absolute value deviates slightly
from the ED and the T-DMRG results. This problem arises since the
maximum in the specific heat occurs at fairly low values of
 $T/J\approx 0.3$. In the purely dimerised case where we have reached
18$^{\rm th}$ order in $\beta$, the HTSE describes the maximum perfectly.
Only in the regime $T/J<0.25$ there is a slight difference to the ED
and to the T-DMRG results. The same is true for the uniform chain
(see inset of Fig.~\ref{fig:heat_comp}) where some deviations from the
exact solution occur well below $T_{\rm max}$. We attribute these
deviations below the maximum position  $T_{\rm max}$ to logarithmic
corrections which cannot be assessed by the high temperature series
expansion.

\begin{figure}[htb]
  \begin{center}
    \leavevmode
    \includegraphics[angle=-90,width=\columnwidth]{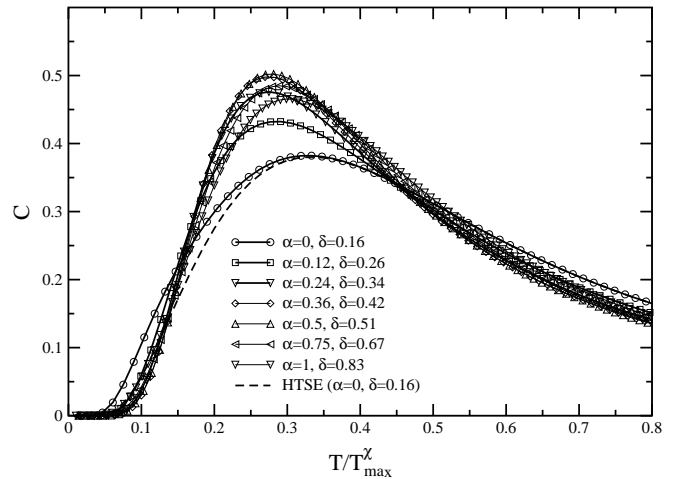}
    \caption{Re-scaled specific heat $C$ for various values of $\delta$ and
      $\alpha$ all yielding the same value $\chi_{\rm max}T^\chi_{\rm max}=0.15$ emu K/mol.}
    \label{fig:heat_distinguish}
  \end{center}
\end{figure}
Fig.~\ref{fig:heat_distinguish} displays $C(T)$ for the points in
Fig.~\ref{fig4} belonging to $\chi_{\rm max}T^\chi_{\rm max}=0.15$
emu K/mol. Therefore, the temperature dependence is given in units
of the maximum temperature $T^\chi_{\rm max}$ of the
susceptibility. Clearly, the curves differ from each other. Hence
the knowledge of $C(T)$, in addition to the knowledge of
$\chi(T)$, renders a complete determination of all three couplings
possible. This is our main point in the present section. In other
words, the knowledge of $\chi(T)$ allows to fix $\delta$ and $J$
for given $\alpha$. But $\alpha$ cannot be determined easily since
there are sets of parameters leading to very similar $\chi(T)$
curves, see Fig.~\ref{fig3}.  The corresponding $C(T)$ curves,
however, differ significantly as illustrated in
Fig.~\ref{fig:heat_distinguish} and thus provide a proper distinction of
different parameter sets.

\begin{figure}[htb]
  \begin{center}
    \leavevmode
    \includegraphics[angle=-90,width=\columnwidth]{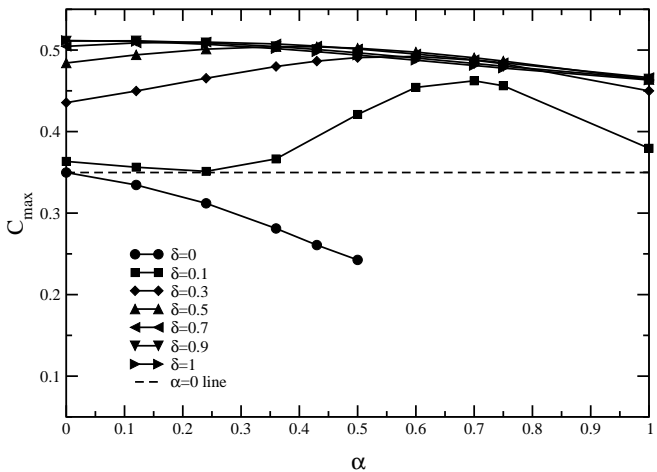}
    \caption{Maximum value $C_{\rm max}$ 
      of the specific heat as function of the
      frustration $\alpha$ for various values of the dimerisation $\delta$.
      For $\delta=0$ no data is shown beyond $\alpha=0.5$ since the exact
      diagonalisation results  are not reliable there.
      }
    \label{fig:c1}
  \end{center}
\end{figure}

\begin{figure}[htb]
  \begin{center}
    \leavevmode
    \includegraphics[angle=-90,width=\columnwidth]{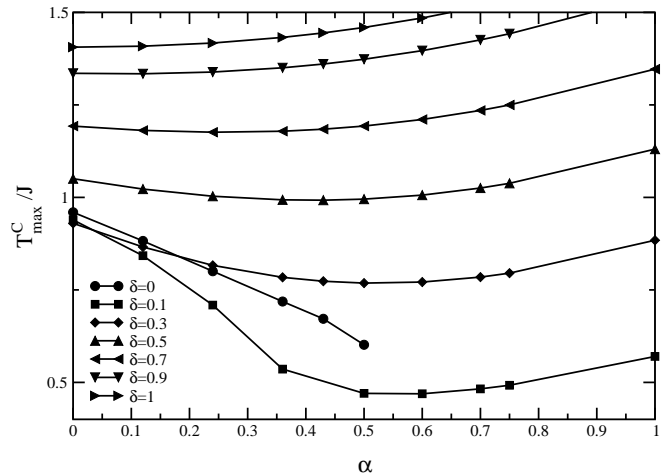}
    \caption{Position $T_{\rm max}$ of the maximum of $C(T)$  as function of
      the frustration $\alpha$ for various values of the dimerisation
      $\delta$. For $\delta=0$ no data is shown beyond $\alpha=0.5$ since 
      the exact diagonalisation results  are not reliable there.}
    \label{fig:c2}
  \end{center}
\end{figure}

To complete our analysis for the specific heat we provide for
$C(T)$ in the Figs.~\ref{fig:c1} and \ref{fig:c2} the analoga of
Figs.~\ref{fig1} and \ref{fig2} for $\chi(T)$. Fig.~\ref{fig:c1}
displays the dimensionless (if $k_{\rm B}$ is set to unity)
specific heat which is independent of the value of the exchange
coupling $J$. For given dimerisation $\delta$ the frustration
parameter $\alpha$ can be read off. Once $\alpha$ is known the
curves in Fig.~\ref{fig:c2} allow to determine the energy scale.

Let us briefly describe the principal behaviour of $C(T)$ as a
function of $\delta$ and $\alpha$. Some of the features of the
curves can be understood by simple arguments. In Fig.~\ref{fig:c1}
all curves converge to the dotted line for increasing values of
$\alpha$. The value of the dotted line is $C_{\rm max}$ of a
uniform chain without dimerisation and frustration. This is
implied by the simple fact that the system approaches the limit of
two independent chains for $\alpha\to\infty$. Then the couplings
$J(1+\delta)$ and $J(1-\delta)$ between the two legs (cf.\
Fig.~\ref{fig:dimfrustHM}) become less and less important
\begin{equation}
\lim_{\alpha \to \infty}C_{\rm max}(\alpha,\delta) =C_{\rm
max}(\alpha=0,\delta=0) \ . \label{eq:limit1}
\end{equation}

For fixed value of the dimerisation $\delta$ the position $T_{\rm
max}$ of the maximum of $C$ is shifted to lower values on
increasing $\alpha$. This can be understood by the suppression of
the dispersion of the elementary excitations due to the
frustration \cite{uhrig96b,uhrig96be,knett00a}.  Thereby the
overall energy scale on which excitations exist is reduced. The same was 
observed for the
susceptibility as shown in the preceding section. A
minimum is reached for a certain value of $\alpha\approx 0.5$
because $T_{\rm max}$ has to rise again since the system
approaches the limit of two independent chains. Quantitatively,
the large $\alpha$ limit fulfills
\begin{equation}
\lim_{\alpha \to \infty}\frac{T_{\rm max}(\alpha,\delta)}{J
\alpha} =\frac{T_{\rm max}(\alpha=0,\delta)}{J} \ .
\label{eq:limit2}
\end{equation}
independent  of $\delta$.

It should be noted that for large $\delta$ (cf.~Fig.~\ref{fig:c2})
the value of $C_{\rm max}$ does not substantially change which
makes it difficult to discriminate the curves experimentally. The
data shown are obtained for a $N=16$ cluster. A finite size
analysis shows that the finite cluster results coincide with the
infinite chain result except for points with $\delta=0$ and
$\alpha>0.5$.

\section{Summary}
\label{sec:Summary} The aim of the present article was two-fold.
In the first place, we provided results and tools to facilitate
and to expedite the analysis of experimental data in terms of a
one-dimensional $S=1/2$ model, namely dimerised and frustrated
spin chains. This model can also be seen as zig-zag chain and
comprises in particular the usual spin ladder. Secondly, we
demonstrated to which extent it is possible to determine the model
parameters quantitatively from the temperature dependences of the
magnetic susceptibility $\chi$ and of the specific heat $C$.

We showed in detail how analytic high temperature series in
high orders can be used to obtain reliable approximants, namely
Dlog-Pad´e approximants. The key point is to use additional
well-known information on the $T=0$ and on the low-temperature
behaviour to stabilise the approximants in the low-temperature
region. We use the size of the gap, the form (linear or quadratic)
of the dispersion in the vicinity of its minimum and the
dimensionality of the system as additional input. Thereby we
achieve very good results in a straightforward fashion. The
validity of the results is comparable to the one achieved by
another extrapolation procedure introduced previously \cite{buhle00}.
The approach used in the present work is simpler since it requires
less additional input. In Ref.~\onlinecite{buhle00} knowledge of the
whole dispersion was used.

With the help of a computer algebra programme the approximants can
be computed very quickly and easily. Thereby efficient data
analysis becomes possible.

The extrapolated series expansion results were gauged carefully by
comparing them to numerical data. The methods employed are exact
diagonalisation, quantum Monte Carlo and temperature
density-matrix renormalisation.

To ease data analysis further we included in the present work
results for many sets of parameters. Figs.~\ref{fig1}, \ref{fig2},
\ref{fig:c1} and \ref{fig:c2} make it possible to read off the
coupling parameters $J, \alpha$ and $\delta$ if as little as the
maximum values of the magnetic susceptibility, of the specific
heat and the corresponding positions $T_{\rm max}^\chi$ and
$T_{\rm max}^C$ are known.

It turned out that the knowledge of $\chi(T)$ at moderate and high
temperatures alone is {\em not} sufficient to determine the three
model parameters. Any additional knowledge, for instance on $C(T)$
or on the singlet-triplet gap $\Delta$, solves the problem. But
such additional information is difficult to obtain. The specific
heat is mostly dominated by the phonon contribution making it
difficult to be extracted. The gap $\Delta$ is in principle well
defined. Frequently, however, the real systems lose their
one-dimensionality at low energies, for instance due to small
interchain couplings. Then the gap is influenced decisively by
these additional residual couplings although the behaviour at
moderate and higher temperatures is perfectly described by a
one-dimensional model. In the analysis of experimental data it is
certainly helpful to consider these facts.

\begin{acknowledgments}
We wish to acknowledge E.~M\"uller-Hartmann for useful discussions
and support. We thank R.~Raupach and F.~Sch\"onfeld for placing
their T-DMRG programme at our disposal, C.~Knetter for help in computing
the gaps from his series and A.~Kl\"umper for providing
the data for the undimerised, unfrustrated spin chain.
This work is supported by the DFG in the Schwerpunkt 1073.
\end{acknowledgments}

%\bibliographystyle{prsty}
%\bibliography{../../bibinput/liter10}

\clearpage

\begin{widetext}

\appendix

\section{Dimerised and frustrated chain}
\label{sec:dimfrustcoeff}

\subsection{Specific heat}

\renewcommand{\arraystretch}{1.5} % zus"atzlicher Zeilenabstand
\renewcommand{\arrayrulewidth}{0.4pt} %Liniendicke in Tabellen
\renewcommand{\doublerulesep}{0pt} % Linienabstand in Tabellen

\begin{table}[hb]
\caption{
Series coefficients $a_{n,k,l}$ for the high temperature expansion
of the  magnetic specific heat of the dimerised, frustrated chain $C =
  \sum_{n,k,l}a_{n,k,l} \alpha^k\delta^l(\beta J)^n$.
Only nonzero coefficients are presented.}
%\begin{table}[h]
\begin{ruledtabular}
\begin{tabular}{c >{$}c <{$} c >{$}c <{$} c >{$}c <{$} %
    c >{$}c <{$} c >{$}c <{$} c >{$}c <{$}}
  (n,k,l) & a_{n,k,l} &  (n,k,l) & a_{n,k,l} &
  (n,k,l) & a_{n,k,l} &  (n,k,l) & a_{n,k,l} &
  (n,k,l) & a_{n,k,l} &  (n,k,l) & a_{n,k,l}   \\ \hline \hline \hline
  % row 1
  (2,0,0)& \frac{3}{16 }&
  (6,0,0)& \frac{21}{4096 }&
  (7,2,4)& \frac{-819}{20480 }&
  (8,4,0)& \frac{35}{2048 }&
  (9,3,6)& \frac{1131}{35840 }&
  (10,2,4)& \frac{-276889}{5505024 }\\
  % row 2
  (2,0,2)& \frac{3}{16 }&
  (6,0,2)& \frac{-177}{4096 }&
  (7,3,0)& \frac{-413}{4096 }&
  (8,4,2)& \frac{-259}{2048 }&
  (9,4,0)& \frac{-261}{286720 }&
  (10,2,6)& \frac{51407}{917504 }\\
  % row 3
  (2,2,0)& \frac{3}{16 }&
  (6,0,4)& \frac{-333}{4096 }&
  (7,3,2)& \frac{651}{10240 }&
  (8,4,4)& \frac{1287}{10240 }&
  (9,4,2)& \frac{-111}{896 }&
  (10,2,8)& \frac{-214825}{3670016 }\\ \cline{1-2}
  % row 4
  (3,0,0)& \frac{3}{32 }&
  (6,0,6)& \frac{73}{4096 }&
  (7,3,4)& \frac{1379}{20480 }&
  (8,5,0)& \frac{-407}{61440 }&
  (9,4,4)& \frac{21597}{286720 }&
  (10,3,0)& \frac{59305}{2752512} \\
  % row 5
  (3,0,2)& \frac{9}{32 }&
  (6,1,0)& \frac{63}{512 }&
  (7,4,0)& \frac{651}{20480 }&
  (8,5,2)& \frac{2951}{61440 }&
  (9,5,0)& \frac{5901}{81920 }&
  (10,3,2)& \frac{38011}{393216 }\\

  % row 6
  (3,1,0)& \frac{-9}{32 }&
  (6,1,2)& \frac{15}{128 }&
  (7,4,2)& \frac{637}{20480 }&
  (8,6,0)& \frac{-2449}{40960 }&
  (9,5,2)& \frac{-26673}{286720 }&
  (10,3,4)& \frac{-297061}{2752512 }\\
  % row 7
  (3,1,2)& \frac{9}{32 }&
  (6,1,4)& \frac{-123}{512 }&
  (7,5,0)& \frac{-245}{8192 }&
  (8,6,2)& \frac{6943}{122880 }&
  (9,5,4)& \frac{537}{81920 }&
  (10,3,6)& \frac{-35787}{917504 }\\
  % row 8
  (3,3,0)& \frac{3}{32 }&
  (6,2,0)& \frac{-363}{4096 }&
  (7,5,2)& \frac{3297}{40960 }&
  (8,8,0)& \frac{1417}{327680 }&
  (9,6,0)& \frac{-2411}{143360 }&
  (10,4,0)& \frac{-138811}{2752512 }\\ \cline{1-2} \cline{7-8}
  % row 9
  (4,0,0)& \frac{-15}{256 }&
  (6,2,2)& \frac{303}{2048 }&
  (7,7,0)& \frac{917}{40960 }&
  (9,0,0)& \frac{-4303}{688128 }&
  (9,6,2)& \frac{-27}{143360 }&
  (10,4,2)& \frac{116115}{1835008 }\\ \cline{5-6}
  % row 10
  (4,0,2)& \frac{3}{128 }&
  (6,2,4)& \frac{-579}{4096 }&
  (8,0,0)& \frac{1417}{327680 }&
  (9,0,2)& \frac{-1401}{286720 }&
  (9,7,0)& \frac{-2229}{573440 }&
  (10,4,4)& \frac{5375}{688128 }\\
  % row 11
  (4,0,4)& \frac{-15}{256 }&
  (6,3,0)& \frac{17}{512 }&
  (8,0,2)& \frac{2199}{81920 }&
  (9,0,4)& \frac{3177}{573440 }&
  (9,7,2)& \frac{-11103}{573440 }&
  (10,4,6)& \frac{-125561}{1835008 }\\
  % row 12
  (4,1,0)& \frac{-3}{32 }&
  (6,3,2)& \frac{-45}{512 }&
  (8,0,4)& \frac{5803}{163840 }&
  (9,0,6)& \frac{3531}{286720 }&
  (9,9,0)& \frac{-4303}{688128 }&
  (10,5,0)& \frac{51701}{1376256 }\\ \cline{9-10}
  % row 13
  (4,1,2)& \frac{3}{32 }&
  (6,4,0)& \frac{105}{1024 }&
  (8,0,6)& \frac{11221}{245760 }&
  (9,0,8)& \frac{-18369}{1146880 }&
  (10,0,0)& \frac{-334433}{110100480 }&
  (10,5,2)& \frac{-14075}{688128 }\\
  % row 14
  (4,2,0)& \frac{-3}{32 }&
  (6,4,2)& \frac{-69}{1024 }&
  (8,0,8)& \frac{-4997}{983040 }&
  (9,1,0)& \frac{2613}{573440 }&
  (10,0,2)& \frac{-249061}{22020096 }&
  (10,5,4)& \frac{-83087}{1376256 }\\
  % row 15
  (4,4,0)& \frac{-15}{256 }&
  (6,6,0)& \frac{21}{4096 }&
  (8,1,0)& \frac{-4793}{61440 }&
  (9,1,2)& \frac{-7821}{143360 }&
  (10,0,4)& \frac{-133325}{11010048 }&
  (10,6,0)& \frac{27641}{2752512 }\\ \cline{1-2}\cline{3-4}
  % row 16
  (5,0,0)& \frac{-15}{256 }&
  (7,0,0)& \frac{917}{40960 }&
  (8,1,2)& \frac{-1999}{20480 }&
  (9,1,4)& \frac{-5913}{286720 }&
  (10,0,6)& \frac{-33811}{3670016 }&
  (10,6,2)& \frac{3069}{57344 }\\
  % row 17
  (5,0,2)& \frac{-15}{128 }&
  (7,0,2)& \frac{1393}{40960 }&
  (8,1,4)& \frac{59}{4096 }&
  (9,1,6)& \frac{14211}{143360 }&
  (10,0,8)& \frac{-431449}{22020096 }&
  (10,6,4)& \frac{-9087}{131072 }\\
  % row 18
  (5,0,4)& \frac{-35}{256 }&
  (7,0,4)& \frac{623}{40960 }&
  (8,1,6)& \frac{1981}{12288 }&
  (9,1,8)& \frac{-16347}{573440 }&
  (10,0,10)& \frac{49649}{36700160 }&
  (10,7,0)& \frac{-1817}{917504 }\\
  % row 19
  (5,1,0)& \frac{25}{128 }&
  (7,0,6)& \frac{399}{8192 }&
  (8,2,0)& \frac{2323}{24576 }&
  (9,2,0)& \frac{3855}{57344 }&
  (10,1,0)& \frac{92629}{2752512 }&
  (10,7,2)& \frac{-53813}{2752512 }\\
  % row 20
  (5,1,4)& \frac{-25}{128 }&
  (7,1,0)& \frac{-2611}{40960 }&
  (8,2,2)& \frac{-1347}{40960 }&
  (9,2,2)& \frac{-6261}{286720 }&
  (10,1,2)& \frac{6463}{172032 }&
  (10,8,0)& \frac{38993}{1572864 }\\
  % row 21
  (5,2,0)& \frac{-5}{128 }&
  (7,1,2)& \frac{1281}{40960 }&
  (8,2,4)& \frac{-1369}{40960 }&
  (9,2,4)& \frac{-7923}{57344 }&
  (10,1,4)& \frac{-981}{458752 }&
  (10,8,2)& \frac{-324557}{11010048 }\\
  % row 22
  (5,2,2)& \frac{-15}{128 }&
  (7,1,4)& \frac{-2009}{40960 }&
  (8,2,6)& \frac{4679}{40960 }&
  (9,2,6)& \frac{40137}{286720 }&
  (10,1,6)& \frac{621}{114688 }&
  (10,10,0)& \frac{-334433}{110100480 }\\
  % row 23
  (5,3,0)& \frac{15}{128 }&
  (7,1,6)& \frac{3339}{40960 }&
  (8,3,0)& \frac{-59}{960 }&
  (9,3,0)& \frac{1}{10240 }&
  (10,1,8)& \frac{-205055}{2752512 }&
  &\\
  % row 24
  (5,3,2)& \frac{-25}{128 }&
  (7,2,0)& \frac{-119}{4096 }&
  (8,3,2)& \frac{-41}{1280 }&
  (9,3,2)& \frac{1359}{35840 }&
  (10,2,0)& \frac{-420475}{11010048 }&
  &\\
  % row 25
  (5,5,0)& \frac{-15}{256 }&
  (7,2,2)& \frac{1491}{10240 }&
  (8,3,4)& \frac{571}{3840 }&
  (9,3,4)& \frac{-4311}{71680 }&
  (10,2,2)& \frac{6235}{393216 }&
  & \\ \cline{1-2}
\end{tabular}
\end{ruledtabular}
\end{table}

\clearpage

\subsection{Susceptibility}

\renewcommand{\arraystretch}{1.5} % zus"atzlicher Zeilenabstand
\renewcommand{\arrayrulewidth}{0.4pt} %Liniendicke in Tabellen
\renewcommand{\doublerulesep}{0pt} % Linienabstand in Tabellen

\begin{table}[hb]
\caption{
Series coefficients $a_{n,k,l}$ for the high temperature expansion
of the  magnetic susceptibility of the dimerised, frustrated chain
  $\chi =\frac{1}{T}\sum_{n,k,l}a_{n,k,l}\alpha^k\delta^l(\beta J)^n$.
  Only nonzero coefficients are presented.}
\begin{ruledtabular}
\begin{tabular}{c >{$}c <{$} c >{$}c <{$} c >{$}c <{$} %
    c >{$}c <{$} c >{$}c <{$} c >{$}c <{$}}
  (n,k,l) & a_{n,k,l} &  (n,k,l) & a_{n,k,l} &
  (n,k,l) & a_{n,k,l} &  (n,k,l) & a_{n,k,l} &
  (n,k,l) & a_{n,k,l} &  (n,k,l) & a_{n,k,l}   \\ \hline \hline \hline
  % row 1
  (0,0,0)& \frac{1}{4 }&
  (5,3,0)& \frac{-1}{128 }&
  (7,1,6)& \frac{-3167}{1474560 }&
  (8,3,2)& \frac{479}{1290240 }&
  (9,3,4)& \frac{-74989}{27525120 }&
  (10,2,4)& \frac{2661047}{2972712960 }\\\cline{1-2}
  % row 2
  (1,0,0)& \frac{-1}{8 }&
  (5,3,2)& \frac{-1}{384 }&
  (7,2,0)& \frac{-805}{73728 }&
  (8,3,4)& \frac{-3947}{5160960 }&
  (9,3,6)& \frac{138421}{82575360 }&
  (10,2,6)& \frac{-33937}{27525120 }\\
  % row 3
  (1,1,0)& \frac{-1}{8 }&
  (5,4,0)& \frac{1}{512 }&
  (7,2,2)& \frac{-131}{36864 }&
  (8,4,0)& \frac{-59}{20160 }&
  (9,4,0)& \frac{317}{229376 }&
  (10,2,8)& \frac{1754671}{5945425920 }\\\cline{1-2}
  % row 4
  (2,0,2)& \frac{-1}{16 }&
  (5,5,0)& \frac{-7}{5120 }&
  (7,2,4)& \frac{-857}{368640 }&
  (8,4,2)& \frac{5119}{2580480 }&
  (9,4,2)& \frac{4099}{2949120 }&
  (10,3,0)& \frac{-311903}{82575360 }\\\cline{3-4}
  % row 5
  (2,1,0)& \frac{1}{8 }&
  (6,0,0)& \frac{-133}{122880 }&
  (7,3,0)& \frac{3023}{737280 }&
  (8,4,4)& \frac{29}{15360 }&
  (9,4,4)& \frac{12337}{20643840 }&
  (10,3,2)& \frac{-1295087}{743178240 }\\ \cline{1-2}
  % row 6
  (3,0,0)& \frac{1}{96 }&
  (6,0,2)& \frac{-83}{40960 }&
  (7,3,2)& \frac{1009}{122880 }&
  (8,5,0)& \frac{-877}{1290240 }&
  (9,5,0)& \frac{-969}{655360 }&
  (10,3,4)& \frac{-1311053}{743178240 }\\
  % row 7
  (3,1,0)& \frac{1}{128 }&
  (6,0,4)& \frac{-21}{40960 }&
  (7,3,4)& \frac{-583}{245760 }&
  (8,5,2)& \frac{-61}{64512 }&
  (9,5,2)& \frac{-1387}{430080 }&
  (10,3,6)& \frac{943507}{743178240 }\\
  % row 8
  (3,1,2)& \frac{3}{128 }&
  (6,0,6)& \frac{-1129}{368640 }&
  (7,4,0)& \frac{-381}{81920 }&
  (8,6,0)& \frac{5389}{20643840 }&
  (9,5,4)& \frac{69103}{41287680 }&
  (10,4,0)& \frac{9659}{3932160 }\\
  % row 9
  (3,2,0)& \frac{-1}{32 }&
  (6,1,0)& \frac{9}{1280 }&
  (7,4,2)& \frac{113}{147456 }&
  (8,6,2)& \frac{6095}{4128768 }&
  (9,6,0)& \frac{93463}{61931520 }&
  (10,4,2)& \frac{3402433}{1486356480 }\\
  % row 10
  (3,3,0)& \frac{1}{96 }&
  (6,1,2)& \frac{1}{1536 }&
  (7,5,0)& \frac{943}{368640 }&
  (8,7,0)& \frac{-1271}{1720320 }&
  (9,6,2)& \frac{-13529}{20643840 }&
  (10,4,4)& \frac{-128473}{41287680 }\\ \cline{1-2}
  % row 11
  (4,0,0)& \frac{5}{1536 }&
  (6,1,4)& \frac{3}{2560 }&
  (7,5,2)& \frac{-199}{368640 }&
  (8,8,0)& \frac{1269}{4587520 }&
  (9,7,0)& \frac{-67097}{82575360 }&
  (10,4,6)& \frac{102007}{495452160 }\\ \cline{7-8}
  % row 12
  (4,0,2)& \frac{7}{768 }&
  (6,2,0)& \frac{221}{61440 }&
  (7,6,0)& \frac{67}{368640 }&
  (9,0,0)& \frac{3737}{74317824 }&
  (9,7,2)& \frac{5233}{9175040 }&
  (10,5,0)& \frac{-1177787}{825753600 }\\
  % row 13
  (4,0,4)& \frac{7}{512 }&
  (6,2,2)& \frac{49}{30720 }&
  (7,7,0)& \frac{1}{16128 }&
  (9,0,2)& \frac{979}{3440640 }&
  (9,8,0)& \frac{-361}{1720320 }&
  (10,5,2)& \frac{52919}{106168320 }\\\cline{5-6}
  % row 14
  (4,1,0)& \frac{-23}{768 }&
  (6,2,4)& \frac{-117}{20480 }&
  (8,0,0)& \frac{1269}{4587520 }&
  (9,0,4)& \frac{481}{1376256 }&
  (9,9,0)& \frac{3737}{74317824 }&
  (10,5,4)& \frac{153863}{495452160 }\\\cline{9-10}
  % row 15
  (4,1,2)& \frac{-3}{256 }&
  (6,3,0)& \frac{-163}{92160 }&
  (8,0,2)& \frac{89}{229376 }&
  (9,0,6)& \frac{263}{1474560 }&
  (10,0,0)& \frac{-339691}{5945425920 }&
  (10,6,0)& \frac{599639}{594542592 }\\
  % row 16
  (4,2,0)& \frac{1}{512 }&
  (6,3,2)& \frac{97}{30720 }&
  (8,0,4)& \frac{1507}{20643840 }&
  (9,0,8)& \frac{15607}{13762560 }&
  (10,0,2)& \frac{-215221}{5945425920 }&
  (10,6,2)& \frac{-370969}{212336640 }\\
  % row 17
  (4,2,2)& \frac{7}{512 }&
  (6,4,0)& \frac{7}{15360 }&
  (8,0,6)& \frac{-10831}{10321920 }&
  (9,1,0)& \frac{-34337}{23592960 }&
  (10,0,4)& \frac{195049}{2972712960 }&
  (10,6,4)& \frac{-816989}{2972712960 }\\
  % row 18
  (4,3,0)& \frac{-1}{96 }&
  (6,4,2)& \frac{-13}{2560 }&
  (8,0,8)& \frac{9623}{13762560 }&
  (9,1,2)& \frac{-3899}{1474560 }&
  (10,0,6)& \frac{145961}{990904320 }&
  (10,7,0)& \frac{791221}{1486356480 }\\
  % row 19
  (4,4,0)& \frac{5}{1536 }&
  (6,5,0)& \frac{23}{7680 }&
  (8,1,0)& \frac{-23629}{20643840 }&
  (9,1,4)& \frac{-2423}{1835008 }&
  (10,0,8)& \frac{1374211}{1981808640 }&
  (10,7,2)& \frac{14941}{99090432 }\\ \cline{1-2}
  % row 20
  (5,0,0)& \frac{-7}{5120 }&
  (6,6,0)& \frac{-133}{122880 }&
  (8,1,2)& \frac{36983}{20643840 }&
  (9,1,6)& \frac{-12323}{6881280 }&
  (10,0,10)& \frac{-4776949}{29727129600 }&
  (10,8,0)& \frac{-367481}{1486356480 }\\ \cline{3-4}
  % row 21
  (5,0,2)& \frac{1}{1536 }&
  (7,0,0)& \frac{1}{16128 }&
  (8,1,4)& \frac{76009}{20643840 }&
  (9,1,8)& \frac{158933}{165150720 }&
  (10,1,0)& \frac{-22843}{1486356480 }&
  (10,8,2)& \frac{-31027}{99090432 }\\
  % row 22
  (5,0,4)& \frac{23}{3072 }&
  (7,0,2)& \frac{-59}{92160 }&
  (8,1,6)& \frac{-1927}{983040 }&
  (9,2,0)& \frac{14125}{4128768 }&
  (10,1,2)& \frac{-322247}{247726080 }&
  (10,9,0)& \frac{22433}{148635648 }\\
  % row 23
  (5,1,0)& \frac{-49}{6144 }&
  (7,0,4)& \frac{-11}{9216 }&
  (8,2,0)& \frac{-58651}{13762560 }&
  (9,2,2)& \frac{79}{458752 }&
  (10,1,4)& \frac{-15173}{7741440 }&
  (10,10,0)& \frac{-339691}{5945425920 }\\
  % row 24
  (5,1,2)& \frac{-67}{3072 }&
  (7,0,6)& \frac{-307}{92160 }&
  (8,2,2)& \frac{-4203}{655360 }&
  (9,2,4)& \frac{699}{458752 }&
  (10,1,6)& \frac{-14683}{9175040 }&
  &\\
  % row 25
  (5,1,4)& \frac{7}{6144 }&
  (7,1,0)& \frac{5863}{1474560 }&
  (8,2,4)& \frac{1061}{393216 }&
  (9,2,6)& \frac{22877}{20643840 }&
  (10,1,8)& \frac{97039}{70778880 }&
  &\\
  % row 26
  (5,2,0)& \frac{37}{1536 }&
  (7,1,2)& \frac{4289}{491520 }&
  (8,2,6)& \frac{33017}{41287680 }&
  (9,3,0)& \frac{-1249}{35389440 }&
  (10,2,0)& \frac{15205963}{5945425920 }&
  &\\
  % row 27
  (5,2,2)& \frac{1}{512 }&
  (7,1,4)& \frac{571}{98304 }&
  (8,3,0)& \frac{28751}{5160960 }&
  (9,3,2)& \frac{-29209}{9175040 }&
  (10,2,2)& \frac{515117}{123863040 }&
  &\\
\end{tabular}
\end{ruledtabular}
\end{table}

\clearpage

\section{Dimerised chain}
\label{sec:dimcoeff}

\subsection{Specific heat}
 
\renewcommand{\arraystretch}{1.5} % zus"atzlicher Zeilenabstand
\renewcommand{\arrayrulewidth}{0.4pt} %Liniendicke in Tabellen
\renewcommand{\doublerulesep}{0pt} % Linienabstand in Tabellen

\begin{table}[h]
\caption{Series coefficients $a_{n,l}$ for the high temperature
expansion of the  magnetic specific heat of the dimerised chain
  $C = \sum_{n,l}a_{n,l}\delta^l(\beta J)^n$.
  Only nonzero coefficients for orders $n>10$ in $\beta$ are presented.}
\begin{ruledtabular}
\begin{tabular}{c >{$}c <{$} c >{$}c <{$} c >{$}c <{$} %
    c >{$}c <{$} c >{$}c <{$}}
  (n,l) & a_{n,l} &  (n,l) & a_{n,l} &
  (n,l) & a_{n,l} &  (n,l) & a_{n,l} &
  (n,l) & a_{n,l}  \\ \hline \hline \hline
  % row 1
  (11,0)& \frac{37543}{31457280 }&
  (13,0)& \frac{-1925339}{41523609600 }&
  (14,12)& \frac{-1410199439}{543581798400 }&
  (16,6)& \frac{-577431395917}{846417258086400 }&
  (17,14)& \frac{28645978566427}{28566582460416000 }\\
  % row 2
  (11,2)& \frac{-2943281}{1981808640 }&
  (13,2)& \frac{60223033}{39636172800 }&
  (14,14)& \frac{3603293663}{41855798476800 }&
  (16,8)& \frac{-199750102373}{307788093849600 }&
  (17,16)& \frac{-25549233744557}{228532659683328000 }\\ \cline{5-6}
  \cline{9-10}
  % row 3
  (11,4)& \frac{-1231703}{198180864 }&
  (13,4)& \frac{2856603451}{871995801600 }&
  (15,0)& \frac{-31504270817}{362750253465600 }&
  (16,10)& \frac{-141175079351}{217650152079360 }&
  (18,0)& \frac{-80067486241427}{4875363406577664000 }\\
  % row 4
  (11,6)& \frac{-2446939}{330301440 }&
  (13,6)& \frac{63260587}{19818086400 }&
  (15,2)& \frac{-7632645211}{10364292956160 }&
  (16,12)& \frac{-6744074943121}{5078503548518400 }&
  (18,2)& \frac{5145779287709}{108341409035059200 }\\
  % row 5
  (11,8)& \frac{-4199591}{396361728 }&
  (13,8)& \frac{168008737}{58133053440 }&
  (15,4)& \frac{-89314894561}{72550050693120 }&
  (16,14)& \frac{718080430229}{846417258086400 }&
  (18,4)& \frac{8816140320217}{36934571261952000 }\\
  % row 6
  (11,10)& \frac{9792739}{1981808640 }&
  (13,10)& \frac{2484478763}{435997900800 }&
  (15,6)& \frac{-210219285347}{217650152079360 }&
  (16,16)& \frac{-1288308081349}{60942042582220800 }&
  (18,6)& \frac{129505647760939}{406280283881472000 }\\ \cline{1-2}
  \cline{7-8}
  % row 7
  (12,0)& \frac{3987607}{3170893824 }&
  (13,12)& \frac{-140182523}{96888422400 }&
  (15,8)& \frac{-3750596387}{5580773130240 }&
  (17,0)& \frac{184265505341}{3627502534656000 }&
  (18,8)& \frac{211128422793049}{812560567762944000 }\\ \cline{3-4}
  % row 8
  (12,2)& \frac{47342317}{13212057600 }&
  (14,0)& \frac{-369233453}{930128855040 }&
  (15,10)& \frac{-28368692533}{51821464780800 }&
  (17,2)& \frac{501069785641}{1904438830694400 }&
  (18,10)& \frac{19613319318773}{90284507529216000 }\\
  % row 9
  (12,4)& \frac{70948027}{26424115200 }&
  (14,2)& \frac{-967322681}{1195879956480 }&
  (15,12)& \frac{-110329916941}{43530030415872 }&
  (17,4)& \frac{1750455145427}{5193924083712000 }&
  (18,12)& \frac{88360795513999}{406280283881472000 }\\
  % row 10
  (12,6)& \frac{16985329}{19818086400 }&
  (14,4)& \frac{29175427}{1993133260800 }&
  (15,14)& \frac{5917497137}{14510010138624 }&
  (17,6)& \frac{65018005163}{453437816832000 }&
  (18,14)& \frac{5854118848751}{8291434364928000 }\\ \cline{5-6}
  % row 11
  (12,8)& \frac{-238381}{1056964608 }&
  (14,6)& \frac{5006543507}{5979399782400 }&
  (16,0)& \frac{851758334701}{8706006083174400 }&
  (17,8)& \frac{747694294049}{114266329841664000 }&
  (18,16)& \frac{-85986647698741}{325024227105177600 }\\
  % row 12
  (12,10)& \frac{2809583}{377487360 }&
  (14,8)& \frac{2068244921}{1993133260800 }&
  (16,2)& \frac{48804050567}{692523211161600 }&
  (17,10)& \frac{-170193679963}{5713316492083200 }&
  (18,18)& \frac{6827887220393}{1329644565430272000 }\\
  % row 13
  (12,12)& \frac{-40555}{117440512 }&
  (14,10)& \frac{1193462617}{664377753600 }&
  (16,4)& \frac{-448010792927}{1171962357350400 }&
  (17,12)& \frac{-221941384979}{1269625887129600 }&
  &\\
\end{tabular}
\end{ruledtabular}
\end{table}

\clearpage

\subsection{Susceptibility}

\renewcommand{\arraystretch}{1.5} % zus"atzlicher Zeilenabstand
\renewcommand{\arrayrulewidth}{0.4pt} %Liniendicke in Tabellen
\renewcommand{\doublerulesep}{0pt} % Linienabstand in Tabellen

\begin{table}[hb]
\caption{ Series coefficients $a_{n,l}$ for the high temperature
expansion of the  magnetic susceptibility of the dimerised chain
  $\chi =\frac{1}{T}\sum_{n,l}a_{n,l}\delta^l(\beta J)^n$.
  Only nonzero coefficients for orders $n>10$ in $\beta$ are presented.}
\begin{ruledtabular}
\begin{tabular}{c >{$}c <{$} c >{$}c <{$} c >{$}c <{$} %
    c >{$}c <{$} c >{$}c <{$}}
  (n,l) & a_{n,l} &  (n,l) & a_{n,l} &
  (n,l) & a_{n,l} &  (n,l) & a_{n,l} &
  (n,l) & a_{n,l}  \\ \hline \hline \hline
  % row 1
  (11,0)& \frac{-1428209}{54499737600 }&
  (13,0)& \frac{7045849}{809710387200 }&
  (14,12)& \frac{378600476623}{3264752281190400 }&
  (16,6)& \frac{-381013820701}{85699747381248000 }&
  (17,14)& \frac{-22234161829843}{685597979049984000 }\\
  % row 2
  (11,2)& \frac{-2683}{29030400 }&
  (13,2)& \frac{2790083}{118908518400 }&
  (14,14)& \frac{-656281799}{76947023462400 }&
  (16,8)& \frac{-6536122273267}{2742391916199936000 }&
  (17,16)& \frac{52097662147}{7031774144102400 }\\ \cline{5-6}
  \cline{9-10}
  % row 3
  (11,4)& \frac{-674059}{7431782400 }&
  (13,4)& \frac{9861031}{653996851200 }&
  (15,0)& \frac{-65174099663}{28566582460416000 }&
  (16,10)& \frac{176252096491}{342798989524992000 }&
  (18,0)& \frac{1116582102301823}{4475583607238295552000 }\\
  % row 4
  (11,6)& \frac{-56827}{1238630400 }&
  (13,6)& \frac{16757}{19818086400 }&
  (15,2)& \frac{-49949749}{11719623573504 }&
  (16,12)& \frac{32635888627387}{1371195958099968000 }&
  (18,2)& \frac{33268250832001}{27124749134777548800 }\\
  % row 5
  (11,8)& \frac{98731}{707788800 }&
  (13,8)& \frac{-56229473}{3923981107200 }&
  (15,4)& \frac{373889611}{357082280755200 }&
  (16,14)& \frac{-150602416817}{3808877661388800 }&
  (18,4)& \frac{301983779893}{208944145996185600 }\\
  % row 6
  (11,10)& \frac{-9539}{27525120 }&
  (13,10)& \frac{-9467111}{71345111040 }&
  (15,6)& \frac{196672854197}{34279898952499200 }&
  (16,16)& \frac{10785364513867}{5484783832399872000 }&
  (18,6)& \frac{30860378761391}{74593060120638259200 }\\\cline{1-2}
  \cline{7-8}
  % row 7
  (12,0)& \frac{18710029}{2242274918400 }&
  (13,12)& \frac{2172449}{21799895040 }&
  (15,8)& \frac{1134039913}{173130802790400 }&
  (17,0)& \frac{1228965600979}{2590036809744384000 }&
  (18,8)& \frac{-182357607317269}{745930601206382592000 }\\ \cline{3-4}
  % row 8
  (12,2)& \frac{-35760853}{2615987404800 }&
  (14,0)& \frac{-358847}{3957275492352 }&
  (15,10)& \frac{86433646963}{11426632984166400 }&
  (17,2)& \frac{146506668199}{685597979049984000 }&
  (18,10)& \frac{-114706799537333}{248643533735460864000 }\\
  % row 9
  (12,4)& \frac{-50519537}{1046394961920 }&
  (14,2)& \frac{361148659}{35876398694400 }&
  (15,12)& \frac{125245014907}{1713994947624960 }&
  (17,4)& \frac{-134460640739}{62327089004544000 }&
  (18,12)& \frac{-90442750894411}{74593060120638259200 }\\
  % row 10
  (12,6)& \frac{-18989863}{356725555200 }&
  (14,4)& \frac{9492225643}{466393183027200 }&
  (15,14)& \frac{-104930723893}{3808877661388800 }&
  (17,6)& \frac{-804378832603}{228532659683328000 }&
  (18,14)& \frac{-169097404668739}{10656151445805465600 }\\ \cline{5-6}
  % row 11
  (12,8)& \frac{-212653873}{5231974809600 }&
  (14,6)& \frac{5394474767}{296795661926400 }&
  (16,0)& \frac{-258645079463}{498616712036352000 }&
  (17,8)& \frac{-1051886885537}{342798989524992000 }&
  (18,16)& \frac{6285092004168191}{497287067470921728000 }\\
  % row 12
  (12,10)& \frac{-161397977}{523197480960 }&
  (14,8)& \frac{7897101007}{652950456238080 }&
  (16,2)& \frac{-15175092143}{3767021862912000 }&
  (17,10)& \frac{-535584743809}{228532659683328000 }&
  (18,18)& \frac{-468309667465837}{1032826986285760512000 }\\
  % row 13
  (12,12)& \frac{193615997}{5231974809600 }&
  (14,10)& \frac{-10743048697}{652950456238080 }&
  (16,4)& \frac{-8706451935593}{1371195958099968000 }&
  (17,12)& \frac{344293742347}{228532659683328000 }&
  &\\
\end{tabular}
\end{ruledtabular}
\end{table}

%\enlargethispage{\baselineskip}

\end{widetext}
\end{document}